\documentclass[aps,prd,showpacs,letterpaper,preprintnumbers,twocolumn,longbibliography,superscriptaddress,notitlepage,hyperref,nofootinbib]{revtex4-1}%
\pdfoutput=1
\usepackage[caption=false]{subfig}
\usepackage{graphicx}
\usepackage{bm}
\usepackage{epsf}
\usepackage{rotating}
\usepackage{epsfig,graphics,rotate,color}
\usepackage{wrapfig}
\usepackage{amssymb}
\usepackage{amsmath}
\usepackage{amsfonts}
\usepackage{booktabs}
\usepackage{gensymb}
\usepackage{placeins}
\usepackage[colorlinks=true,citecolor=blue,linkcolor=blue, allcolors=blue]{hyperref}

\newcommand{\mdm}{m_{\rm DM}}
\newcommand{\rdm}{\rho_{\rm DM}}

\newcommand{\be}{\begin{equation}}
\newcommand{\ee}{\end{equation}}

\begin{document}

\preprint{\tt IFIC/18-36}

\title{Flavor of cosmic neutrinos preserved by ultralight dark matter}

\author{Yasaman Farzan}
\email{yasaman@theory.ipm.ac.ir}
\affiliation{School of Physics, Institute for Research in Fundamental Sciences (IPM), P.O.~Box 19395-5531, Tehran, Iran}
\author{Sergio Palomares-Ruiz}
\email{sergiopr@ific.uv.es}
\affiliation{Instituto de F\'{\i}sica Corpuscular (IFIC), CSIC-Universitat de Val\`encia,\\ Apartado de Correos 22085, E-46071 Valencia, Spain}

\begin{abstract}
Within the standard propagation scenario, the flavor ratios of high-energy cosmic neutrinos at neutrino telescopes are expected to be around the democratic benchmark resulting from hadronic sources, $\left( 1 : 1 : 1 \right)_\oplus$. We show how the coupling of neutrinos to an ultralight dark matter complex scalar field would induce an effective neutrino mass that could lead to adiabatic neutrino propagation. This would result in the preservation at the detector of the production flavor composition of neutrinos at sources. This effect could lead to flavor ratios at detectors well outside the range predicted by the standard scenario of averaged oscillations. We also present an electroweak-invariant model that would lead to the required effective interaction between neutrinos and dark matter.
\end{abstract}

\maketitle

{\bf Introduction.---}
Although cosmological observations have determined the contribution of dark matter (DM) to the energy budget of the Universe with an outstanding precision, the nature of the particles making up this component of the Universe is still unknown. In particular, the mass, spin and couplings of DM particles have not been determined yet. A lower bound on the mass ($\mdm$) comes from the de Broglie wavelength of the DM particle, $\lambda_{\rm dB} = 2\pi/(\mdm \, v)$, which is required to be smaller than the size of dwarf galaxies. Ultralight bosonic DM with a mass close to this bound, $\sim 10^{-22} - 10^{-21}$~eV, has gained popularity (see, e.g., Refs.~\cite{Suarez:2013iw, Rindler-Daller:2013zxa, Chavanis:2011zi, Marsh:2015xka, Hui:2016ltb, Lee:2017qve} for reviews), as it can address the small structure problems that the canonical cold DM scenario suffers from~\cite{Du:2016zcv, Mocz:2017wlg, Zhang:2017chj}. Recent studies of rotation curves of nearby galaxies~\cite{Bar:2018acw}, of dwarf galaxies~\cite{Marsh:2015wka, Gonzales-Morales:2016mkl, Urena-Lopez:2017tob}, the comparison between the predictions of hydrodynamical simulations and Lyman-$\alpha$ observations~\cite{Irsic:2017yje, Armengaud:2017nkf, Kobayashi:2017jcf, Nori:2018pka}, and analyses of cosmological data~\cite{Li:2013nal, Bozek:2014uqa, Hlozek:2014lca} have set lower bounds of $\sim 10^{-21}$~eV on $\mdm$.

As long as the de Broglie wavelength, $\lambda_{\rm dB}$, is much larger than the average distance between DM particles ($\sim n_{\rm DM}^{-1/3} = (\mdm/\rdm)^{1/3}$), DM can be described by a classical field oscillating in time with a period given by the Compton wavelength, $\lambda_{\rm C} = 2\pi/\mdm$. It has been shown that a Yukawa coupling between neutrinos and the background ultralight scalar DM ($\phi$) can induce a time varying effective neutrino mass, causing spectacular time modulation effects for solar~\cite{Berlin:2016woy} and long-baseline and reactor neutrinos~\cite{Krnjaic:2017zlz, Brdar:2017kbt}.

In this letter, we consider a derivative interaction between the ultralight complex scalar DM and neutrinos of the form
\begin{equation}
\label{eq:JJ}  
i \, \frac{g_\alpha}{\Lambda^2}(\phi^\dagger \partial_\mu \phi - \phi \, \partial_\mu \phi^*) (\bar{\nu}_\alpha \gamma^\mu \nu_\alpha) ~. 
\end{equation}
As we shall see, this effective term can be obtained by integrating out a new neutral gauge boson coupled to the currents of neutrinos and $\phi$ in an ultraviolet complete electroweak-invariant form. By treating $\phi$ as a classical non-relativistic field, we show that this coupling induces a neutrino mass term proportional to
\begin{equation}
V_\alpha \, \nu_\alpha^\dagger \nu_\alpha ~.
\end{equation}
As long as $\Delta V \gg \Delta m^2/E_\nu$ (with $\Delta V$, $\Delta m^2$ and $E_\nu$, being the difference between two $V_\alpha$, the neutrino mass square difference and the neutrino energy, respectively), this new term would dominate the Hamiltonian and therefore, the time evolution of neutrinos. Taking the coupling to be flavor conserving but flavor non-universal ($V_e \ne V_\mu \ne V_\tau$), the outcome would be the flavor conservation in the propagation of high-energy cosmic neutrinos. The oscillation pattern of lower energy neutrinos, such as solar, long-baseline or supernova neutrinos, would not be affected, though. For those energies, $\Delta m^2/E_\nu \gg \Delta V$, and the standard results are recovered. Note that this energy dependence is a characteristic feature of dimension-three operators.

{\bf Flavor of cosmic neutrinos.---}
The study of the flavor composition has been long recognized as a powerful tool to determine the production mechanism of high-energy astrophysical neutrinos~\cite{Rachen:1998fd, Athar:2000yw, Anchordoqui:2003vc, Kashti:2005qa, Kachelriess:2006fi, Mena:2006eq, Lipari:2007su, Pakvasa:2007dc, Esmaili:2009dz, Choubey:2009jq, Lai:2009ke, Hummer:2010ai, Fu:2012zr, Vissani:2013iga, Xu:2014via, Fu:2014isa, Bustamante:2015waa, Shoemaker:2015qul, Palladino:2018qgi}. 

In the standard scenario, astrophysical neutrinos are produced from the decays of pions and kaons and secondary muons, which are in turn created by hadronic (proton-proton, $pp$) or photohadronic (proton-photon, $p\gamma$) interactions in cosmic accelerators. The flavor composition at the source of the neutrino plus antineutrino flux is (approximately) $\left( \nu_e : \nu_\mu : \nu_\tau \right)_{\rm S} \simeq \left( 1 : 2 : 0 \right)_{\rm S}$ in both cases\footnote{Nevertheless, while in the case of $pp$ processes, the flavor ratios for the separate neutrino or antineutrino fluxes are the same, for $p\gamma$ interactions, the flavor ratios are $\left( \nu_e : \nu_\mu : \nu_\tau \right)_{\rm S} \simeq \left( 1 : 1 : 0 \right)_{\rm S}$ for neutrinos and $\left( {\bar\nu}_e : {\bar\nu}_\mu : {\bar\nu}_\tau \right)_{\rm S} \simeq \left( 0 : 1 : 0 \right)_{\rm S}$ for antineutrinos.}. Given the cosmic distances these neutrinos travel, oscillation probabilities are averaged out~\cite{Learned:1994wg}. As a consequence, for the values of the mixing angles measured in neutrino oscillation experiments~\cite{Esteban:2016qun, deSalas:2017kay, Capozzi:2018ubv}, the resulting ($\nu + \bar{\nu}$) flavor composition at detectors at Earth becomes $\left( \nu_e : \nu_\mu : \nu_\tau \right)_\oplus \simeq \left( 1 : 1 : 1 \right)_\oplus$.

\begin{figure}[t]
	\begin{center}
		\includegraphics[width=0.5\textwidth]{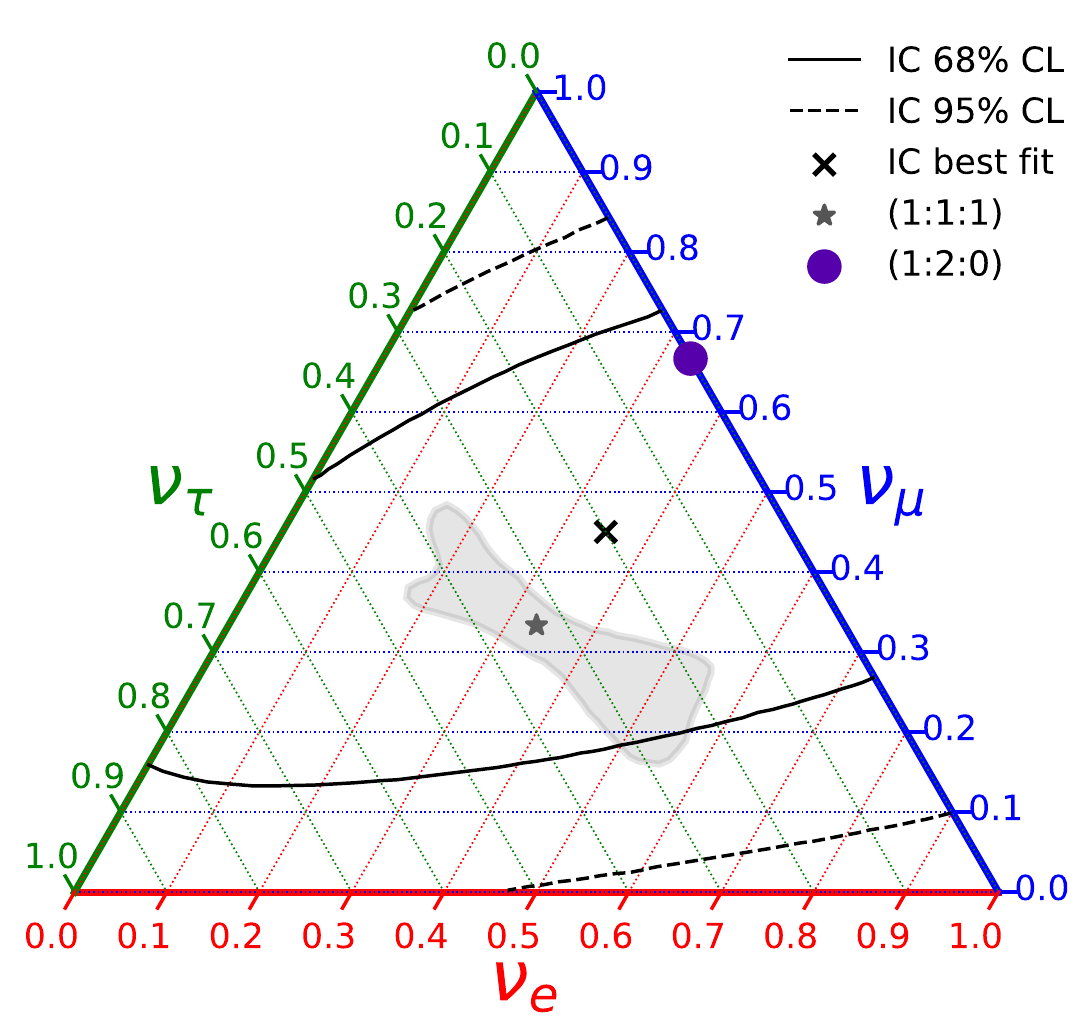}
	\end{center}
	\caption{Ternary plot of the flavor composition of cosmic neutrinos. The allowed flavor compositions are represented by the regions within the black lines, using IceCube HESE events after 7.5~years (68\% and 95\% confidence level), with three types of topologies: muon tracks, single and double cascades~\cite{ICflavorNeutrino18}. Also shown is the obtained best fit (black cross). The gray shaded contour indicates the allowed region after standard averaged oscillations during propagation, and accounts for uncertainties at 95\% confidence level of the neutrino mixing angles~\cite{Esteban:2016qun}. For hadronic sources, the expected flavor ratio at detection after standard propagation lies at the center (star), whereas within the scenario discussed in this letter, it would coincide with the flavor composition at the source (thick purple dot).}
	\label{fig:flavor}
\end{figure}

There are two main features that stand out from the canonical flavor composition. Due to maximal mixing in the $\mu-\tau$ sector, astrophysical $\nu_\mu$ and $\nu_\tau$ fluxes are always expected to be very similar at Earth. Moreover, regardless of  the flavor composition at the cosmic source, all flavors become finally populated after propagation through cosmic distances. Thus, if any of the three neutrino flavors is found not to contribute to the observed high-energy event spectrum in neutrino telescopes, this necessarily implies the existence of new physics. In this paper we present a scenario in which the neutrino flavor composition at the source is preserved and coincides with that at the detector. Given that $\nu_\tau$'s are very scarcely produced at astrophysical sources, this possibility is very far from the canonical expectation. This can be seen in Fig.~\ref{fig:flavor}, where we show the expected flavor combination at Earth from hadronic sources within the scenario discussed in this letter (which coincides with that at production), the current allowed region and the expected flavor composition from standard averaged oscillations.

{\bf Propagation of cosmic neutrinos interacting with ultralight scalar dark matter.---}Now we show how the interaction term in Eq.~(\ref{eq:JJ}) could result in a flavor composition of the cosmic neutrino flux at detection approximately equal to that at the source.

A complex field can be decomposed as
\begin{equation} \label{solution}
\phi (\vec x, \, t) = \psi(\vec x, \, t) \, \frac{1}{\sqrt{2} \, \mdm} \, e^{-i \, \mdm \, t} ~,
\end{equation}
such that, in the non-relativistic limit, $\partial_0 \psi(t) \ll \mdm \, \psi(t)$ and thus, solving the equation of motion (Euler-Lagrange equation), the classical field $\psi$ is approximately constant, and can be obtained from the contribution to the 00 component of the energy-momentum tensor ($T_{00} = \rdm$), which results in $\rdm = |\psi|^2$.

The charge density associated to this complex scalar can be written as
\begin{equation} 
J_0^{\phi} = i \, \left(\phi^* \, \partial_0 \phi - \phi \, \partial_0 \phi^*\right) = \frac{|\psi|^2}{\mdm} = \frac{\rdm}{\mdm} ~,
\end{equation}
which corresponds to the number density of DM particles.

To account for the correction induced on the free Hamiltonian by the derivative interaction, we insert this expression for $J_0^\phi$ in Eq.~(\ref{eq:JJ}), which results in the following effective mass term\footnote{Note that this scenario bears certain similarities with that presented in Refs.~\cite{Ando:2009ts, Klop:2017dim}, with neutrino-dark energy interactions.},
\begin{equation} 
V_\alpha \, \nu_\alpha^\dagger \nu_\alpha = \left(\frac{\rdm}{\mdm}\right) \, \left(\frac{g_\alpha}{\Lambda^2}\right) \, \bar{\nu}_\alpha \gamma^0 \nu_\alpha ~.
\end{equation}
Thus, the total Hamiltonian in the flavor basis for neutrinos propagating in the ultralight scalar DM background is given by
\begin{equation}
H_{\rm DM} = H_{\rm vac} \pm {\rm diag}(V_e, V_\mu, V_\tau) ~, 
\end{equation}
where $H_{\rm vac} = M_\nu^2/ 2 E_\nu$ is the Hamiltonian in vacuum, with $M_\nu^2$ the neutrino mass square matrix in the flavor basis, and the $+$ and $-$ signs correspond to neutrinos and antineutrinos, respectively. Two comments are in order:
\begin{itemize}
	\item The Lorentz structure of the effective mass is similar to the standard matter effects and thus, independent of the neutrino energy. Like the standard MSW effect, the impact of this DM interaction on neutrino propagation becomes more relevant for more energetic neutrinos. Moreover, like standard matter effects, the signs of the effect for neutrinos and antineutrinos are opposite. Thus, it also induces CP (and CPT) violation in neutrino propagation.
	\item This effective matter term, unlike the cases considered in Refs.~\cite{Berlin:2016woy, Krnjaic:2017zlz, Brdar:2017kbt}, does not have a time-dependent oscillatory behavior, but it only depends on the DM density, $\rdm$, along the route of neutrinos.
\end{itemize} 

The local DM density in the solar system is determined to be $\sim 0.3$~GeV/cm$^3$, although with about a factor of two of uncertainty~\cite{Read:2014qva}. On the other hand, a significant fraction of the high-energy neutrino flux is expected to originate at sources which are located at relatively dense parts of the Universe with DM densities that can be orders of magnitude larger than the local value. Analogously to the standard propagation of neutrinos in matter, medium effects become dominant when the potential difference, $\Delta V = (\rdm/\mdm) \, \Delta g/\Lambda^2$, is larger than the difference of their vacuum terms, $\Delta m^2 \, \cos 2\theta/(2 \, E_\nu)$, with $\theta$ being the vacuum mixing angle (within a two-neutrino framework). Thus, if 
\begin{eqnarray}
\frac{\Lambda^2/\Delta g}{(20~{\rm PeV})^2} & \ll  & \left( \frac{E_\nu}{100~{\rm TeV}} \right) \, \left( \frac{10^{-21}~{\rm eV}}{\mdm} \right) \nonumber \\ 
&  \times & \left( \frac{10^{-3}~{\rm eV}^2} {\Delta m^2 \, \cos 2\theta}\right)  \, \left( \frac{\rdm}{0.3~{\rm GeV/cm}^3} \right) ~, 
\label{eq:matter}
\end{eqnarray}
the effective in-medium mixing of neutrinos would be suppressed. That is, the flavor eigenstates would coincide with the mass eigenstates of the effective Hamiltonian. Moreover, if the variation of $H_{\rm DM}$ (determined by $d \rdm/dx$, with $x$ the spatial coordinate) is slow, the evolution would be adiabatic. This implies that an eigenstate of the Hamiltonian at a given point (i.e., the effective mass eigenstate) remains so throughout the propagation, despite the fact that the Hamiltonian changes. Both the sources of high-energy neutrinos and the Earth are located in regions where Eq.~(\ref{eq:matter}) could be satisfied. Therefore, if the propagation is adiabatic, the flavor composition at production would be preserved at the detector. Adiabaticity requires that the variation of the mixing angle in matter, $d\theta_m/dx$, is slow compared to $\Delta m_m^2/(4 \, E_\nu)$, with $\Delta m_m^2$ the mass square difference in the medium. In other words, this occurs if the typical length scale for the variation of the medium density is much larger than the neutrino oscillation length in that medium. The condition for adiabatic propagation is most stringent at resonance (i.e., $2 \, E_\nu \, \Delta V_{\rm res} = \Delta m^2 \, \cos 2\theta$) and therefore, for a two-neutrino system,

\begin{eqnarray}
\left|\frac{d \rdm/dx}{\rdm}\right|_{\rm res} & \ll & \Delta V_{\rm res} \, \tan^2 2\theta \\
& \sim & 5 \times 10^4 \, {\rm pc}^{-1} \, \left( \frac{\sin^2 2\theta}{\cos 2\theta}\right) \nonumber \\
& & \times \left( \frac{\Delta m^2}{7\times 10^{-5}~{\rm eV}^2} \right) \, \left( \frac{100~{\rm TeV}}{E_\nu}\right)_{\rm res} ~, \nonumber 
\end{eqnarray}
where the subindex $_{\rm res}$ indicates quantities evaluated at resonance. Notice that for the energies satisfying Eq.~(\ref{eq:matter}), the adiabaticity condition would be more easily met. Indeed, it is easily satisfied for high-energy neutrinos. The large de Broglie wavelength of ultralight DM ($> 10$~pc) prevents the existence of very sharp features in the DM distribution at galactic scales and thus, $(d\rdm/dx)/\rdm < 0.1~{\rm pc}^{-1}$. Notice that neutrinos on their path to Earth may pass through voids, but as long as their production site is located within a relatively dense region inside a DM halo and Eq.~(\ref{eq:matter}) is satisfied at Earth, their initial flavor composition would be preserved at detection. The differences between this scenario and the standard propagation, and the currently allowed experimental region can be seen in Fig.~\ref{fig:flavor}.

{\bf Example of an underlying model.---} 
In this section, we show an example for building a model which leads to the effective coupling shown in Eq.~(\ref{eq:JJ}). The complex scalar field which is assumed to play the role of DM, $\phi$, is taken to be  charged under a new $U(1)$. Taking the gauge coupling to be $g_\phi$, the DM field would have a coupling with the new gauge boson, $Z^\prime$, of the form: 
\begin{equation} 
i \, g_\phi \, (\phi^* \, \partial^\mu \phi - \phi \, \partial^\mu \phi^*) \, Z^\prime_\mu ~.
\label{eq:phiTOz}
\end{equation} 
Taking the $U(1)$ charges of leptons of generation $\alpha$ to be $g_\alpha$, the left-handed lepton doublets, $L_\alpha$, would then couple to $Z^\prime$ as 
\begin{equation} 
 \, Z^\prime_\mu \, \sum_{\alpha} g_\alpha \, \bar{L}_\alpha \gamma^\mu L_\alpha  ~. 
\label{eq:mu-tau} 
\end{equation}
If $g_e + g_\mu + g_\tau = 0$, the triangle anomalies that involve one new $U(1)$ vertex automatically cancel, with or without adding right-handed neutrinos. If a right-handed neutrino with the same $U(1)$ charge is added for each $\nu_\alpha$, the $U(1)-U(1)-U(1)$ anomaly would cancel, too. However, without right-handed neutrinos, the cancellation of this anomaly requires $g_e^3 + g_\mu^3 + g_\tau^3 = 0$. 

Integrating out $Z^\prime$, the effective coupling in Eq.~(\ref{eq:JJ}) would be given in terms of the $Z'$ mass and $g_\alpha$,
\begin{equation}
\Lambda^2 = \frac{m_{Z'}^2}{g_\phi} ~.
\end{equation}
 Combined with the condition for the matter effects dominance, Eq.~(\ref{eq:matter}), this implies
 \begin{eqnarray}
 \frac{m_{Z'}}{ \sqrt{\Delta g \, g_\phi }} & \ll & 20~{\rm PeV} \left( \frac{E_\nu}{100~{\rm TeV}} \right)^{1/2} \,
  \left( \frac{10^{-21}~{\rm eV}}{\mdm} \right)^{1/2} \nonumber \\ 
 & & \times \left( \frac{10^{-3}~{\rm eV}^2}{\Delta m^2 \, \cos 2\theta} \right)^{1/2} \left( \frac{\rdm}{0.3~{\rm GeV/cm}^3} \right)^{1/2} ~.
 \label{eq:mZ'}
 \end{eqnarray}

For $Z^\prime$ heavier than $\sim$TeV, there is practically no observational bound on the coupling constants and $g_\alpha$ could be as large as ${\cal O}(1)$~\cite{Kaneta:2016vkq}. However, an upper bound on $m_{Z'}$ can be deduced from the theoretical discussion on the stabilization of the DM mass. Like most models with a new unprotected scalar (including the SM Higgs), this model encounters a hierarchy problem, because of the radiative contribution to the scalar mass given by a high-energy scale cutoff. Just like in the case of the SM, we shall assume there is a mechanism (e.g., SUSY-like) that manages to cancel out this contribution. Nevertheless, even after assuming there is some extra mechanism to cancel the cutoff dependent contribution, to avoid fine tuned cancellations, the radiative correction to the $\phi$ mass should not be much larger than $\mdm$. Thus, assuming all the couplings to be of the same order ($g_\phi \sim g_\alpha$, although this is, of course, not necessary), the condition on the radiative contribution to $\mdm$ ($m_\phi \gtrsim m_{Z'} g_\phi/(4\pi)$) implies $m_{Z'} \lesssim 0.01$ eV and $g_\alpha \sim g_\phi \gtrsim 10^{-18}$, for the benchmark parameters used in Eq.~(\ref{eq:mZ'}). For such a light $Z'$, the bounds on $g_e$ from extra long-range interactions are strong~\cite{Adelberger:2006dh}, which can be circumvented by the anomaly-free $\mathcal{L}_\mu - \mathcal{L}_\tau$ gauge symmetry with $g_e = 0$ and $g_\mu = -g_\tau$ and therefore, $V_e = 0$ and $V_\mu = -V_\tau$. Moreover, notice that coherent forward scattering of neutrinos off the DM background would be mediated by a $t$-channel and so, the virtual $Z'$ would carry zero energy-momentum. Thus, despite $E_\nu \gg m_{Z'}$, $Z'$ could be integrated out, leading to the effective Lagrangian in Eq.~(\ref{eq:JJ}).
 
Neutrino mixing then requires breaking of this $U(1)$. This can be accommodated within the seesaw mechanism where three right-handed neutrinos, $N_e$, $N_\mu$ and $N_\tau$ with $U(1)$ charges equal to those of corresponding left-handed leptons, are introduced.  For the special case of $\mathcal{L}_\mu - \mathcal{L}_\tau$, adding new scalars charged under $U(1)$, $S_1$ and $S_2$, couplings of the form $S_1 \overline{N_\mu^c} N_\mu$, $S_1^* \overline{N_\tau^c} N_\tau$, $S_2 \overline{N_e^c} N_\mu$ and $S_2^* \overline{N_e^c} N_\tau$ can be written. The vacuum expectation values (VEVs) of $S_1$ and $S_2$ can lead to mixing between flavors, reproducing the flavor structure of the neutrino mass mixing. Their VEVs also contribute to the $Z^\prime$ mass, as $ g_\mu \, \sqrt{4 \, \langle S_1\rangle^2 + \langle S_2\rangle^2}$. Taking $g_\mu \sim 10^{-18}$ and $\langle S_1\rangle\sim \langle S_2\rangle \sim 20$ PeV, a seesaw mechanism, as well as a naturally small $m_{Z'}$, could be realized.

{\bf Discussion.---} 
The flavor composition of high-energy cosmic neutrinos is a diagnostic tool for different production mechanisms at astrophysical accelerators. By considering a flavor-diagonal interaction between neutrinos and an ultralight scalar DM candidate given by Eq.~(\ref{eq:JJ}), we have shown that the  flavor composition of the neutrino flux at production could be preserved at detection. This would occur if the induced potential (proportional to the DM number density) is larger than the vacuum oscillations term. This result is in clear contrast to the standard picture which predicts complete reshuffling of the flavor composition after propagation. For example, within the standard scenario, the canonical initial composition for hadronic sources $(\nu_e : \nu_\mu : \nu_\tau)_{\rm S} = (1 : 2 : 0)_{\rm S}$ would be converted into $(\nu_e : \nu_\mu : \nu_\tau)_\oplus = (1 : 1 : 1)_\oplus$ at Earth, but with this new interaction the $(\nu_e : \nu_\mu : \nu_\tau)_\oplus = (1 : 2 : 0)_\oplus$ ratio would be preserved. Notice also that, in the presence of the new interaction, neutrinos could decay as $\nu_i \to \nu_j \, \phi \, \phi$ or even $\nu_i \to \nu_j \, Z'$. This would produce an additional modification of the flavor composition detected at Earth. For the small couplings considered here, however, their lifetime would be much longer than the age of the Universe, rendering neutrino decay irrelevant.

One key consequence of the modification of the flavor compositions is related to the detection of tau neutrinos. Direct production of tau neutrinos in cosmic accelerators is highly suppressed and a $\nu_\tau$ and $\bar{\nu}_\tau$ flux at Earth would be created by neutrino mixing and propagation along astrophysical distances. Nevertheless, within the scenario presented here, the neutrino flux at Earth would contain barely no $\nu_\tau$ or $\bar{\nu}_\tau$. Thus, a powerful tool to discriminate it from the standard scenario is to measure the $\nu_\tau$ and $\bar{\nu}_\tau$ content of the high-energy cosmic neutrino flux. 

Indeed, there are two $\nu_\tau$ candidates in the latest IceCube HESE sample~\cite{ICflavorNeutrino18}, in agreement with expectations from standard propagation~\cite{Aartsen:2017mau}. This is why the best fit in Fig.~\ref{fig:flavor} is not along the $\nu_e - \nu_\mu$ side, as happened in previous analyses~\cite{Palomares-Ruiz:2015mka, Aartsen:2015knd, Aartsen:2017mau}. If this is confirmed with future data and improved analyses, this ultralight DM scenario would be disfavored, and a lower bound on $\Lambda^2/\Delta g$ could be set. As of now, and as can be seen from Fig.~\ref{fig:flavor}, the unaltered $(1 : 2 : 0)_\oplus$ flavor composition from hadronic sources is allowed within a 68\% confidence level. The $(0 : 1 : 0)_\oplus$ and $(1 : 0 : 0)_\oplus$ flavor ratios, however, are already ruled out at more than 95\% confidence level. This means that, within our scenario, the sources of the cosmic neutrinos are unlikely to be purely pion decays with a stopped muon or purely neutron decays, respectively.

It is also interesting to compare these effects to already existing limits on similar CPT-violating neutrino interaction terms. Indeed, this effective potential has already been (locally) constrained by IceCube, using atmospheric neutrinos with energies $\lesssim 1$~TeV. The limit on constant couplings of dimension-three operators is $\lesssim 10^{-24}$~GeV~\cite{Aartsen:2017ibm}, as could be expected from the vacuum oscillations term, $\Delta m^2 / (2 \, E_\nu) = 5 \times 10^{-25}~{\rm GeV} \, (\Delta m^2/10^{-3}~{\rm eV}^2) \, (1~{\rm TeV}/E_\nu)$. Note that the effective interaction in the rest frame of the DM background has the Lorentz structure of a mass term, $\nu^\dagger \nu$, so similarly to the standard scenario, the lower the neutrino energy the more suppressed matter effects are. Thus, for values close to the saturation of this bound, matter effects would be the dominant ones at the higher energies considered here, and have no effect on atmospheric neutrinos. Indeed, oscillations of solar, atmospheric, supernova and terrestrial neutrinos would remain unaffected.

Interestingly, the $U(1)$ gauge interaction in Eq.~(\ref{eq:phiTOz}) would also lead to a repulsive force among ultralight scalar DM particles, in contrast to scenarios with axion-like particles, for which self-interactions are usually attractive. The low scale required in our scenario, $\Lambda/\sqrt{g_\alpha} \lesssim {\cal O}(10)$~PeV, would imply the instability of DM structures as light as a solar mass if the force were attractive~\cite{Chavanis:2011zi, Chavanis:2011zm, Eby:2015hsq, Chavanis:2016dab, Eby:2016cnq, Levkov:2016rkk}. Nevertheless, being repulsive, the maximum mass before collapse would be much larger than that of superclusters of galaxies~\cite{Chavanis:2011zi, Chavanis:2011zm, Eby:2015hsq}.

It is only a few years since the first detection of high-energy extraterrestrial neutrino events and most questions about the origin of this neutrino flux are awaiting for answers. The presence of new physics affecting the expected signatures could make more difficult (and interesting) the interpretation of present and future observations. Here, we have investigated a rather speculative scenario that could give rise to striking effects on the high-energy neutrino flavor composition at detection.

{\bf Acknowledgments.}
YF thanks A.~Smirnov and L.~Visenelli for fruitful discussions.
This project has received funding from the European Union's Horizon 2020 research and innovation programme under the Marie Sk\l{}odowska-Curie grant agreement No.~674896 and No.~690575.
YF is also grateful to the ICTP associate office and to IFIC for warm and generous hospitality. The authors thank Iran National Science Foundation (INSF) for partial financial support under contract 94/saad/43287.
SPR is supported by a Ram\'on y Cajal contract, by the Spanish MINECO under grants FPA2017-84543-P, FPA2014-54459-P and SEV-2014-0398 and by the Generalitat Valenciana under grant PROMETEOII/2014/049, and partially, by the Portuguese FCT through the CFTP-FCT Unit 777 (PEst-OE/FIS/UI0777/2013). SPR also acknowledges the IPM for warm and generous hospitality. 

\vspace{-1cm}

\bibliography{fuzzybiblio}

\end{document}